\documentstyle[prd,aps,preprint,tighten,epsfig]{revtex}

\begin{document}

\draft

\title{Deviations from Tri-bimaximal Neutrino Mixing in Type-II
Seesaw and Leptogenesis}
\author{{\bf Aik Hui Chan} $^b$, ~ {\bf Harald Fritzsch} $^c$,
~ {\bf Shu Luo} $^a$
\thanks{E-mail: luoshu@mail.ihep.ac.cn},
~ {\bf Zhi-zhong Xing} $^a$
\thanks{E-mail: xingzz@mail.ihep.ac.cn}}
\address{$^a$ Institute of High Energy Physics, Chinese Academy of
Sciences, Beijing 100049, China  \\
$^b$ Department of Physics, National University of Singapore,
Singapore 117542,
Singapore ~ \\
$^c$ Sektion Physik, Ludwig-Maximilians-Universit$\it\ddot{a}$t
M$\it\ddot{u}$nchen, 80333 Munich, Germany ~~~}

\maketitle

\begin{abstract}
Current experimental data allow the zero value for one neutrino
mass, either $m^{}_1 =0$ or $m^{}_3 =0$. This observation implies
that a realistic neutrino mass texture can be established by
starting from the limit (a) $m^{}_1 = m^{}_2 =0$ and $m^{}_3 \neq
0$ or (b) $m^{}_1 = m^{}_2 \neq 0$ and $m^{}_3 =0$. In both cases,
we may introduce a particular perturbation which ensures the
resultant neutrino mixing matrix to be the tri-bimaximal mixing
pattern or its viable variations with all entries being formed
from small integers and their square roots. We find that it is
natural to incorporate this kind of neutrino mass matrix in the
minimal Type-II seesaw model with only one heavy right-handed
Majorana neutrino $N$ in addition to the $SU(2)^{}_L$ Higgs
triplet $\Delta^{}_L$. We show that it is possible to account for
the cosmological baryon number asymmetry in the $m^{}_3 =0$ case
via thermal leptogenesis, in which the one-loop vertex correction
to $N$ decays is mediated by $\Delta^{}_L$ and the CP-violating
asymmetry of $N$ decays is attributed to the electron flavor.
\end{abstract}

\pacs{PACS number(s): 14.60.Pq, 13.10.+q, 25.30.Pt}

\newpage

\section{Introduction}

Recent solar \cite{SNO}, atmospheric \cite{SK}, reactor \cite{KM}
and accelerator \cite{K2K} neutrino experiments have convincingly
verified the hypothesis of neutrino oscillation, a quantum
phenomenon which can naturally happen if neutrinos are slightly
massive and lepton flavors are not conserved. The mixing of lepton
flavors is described by a $3\times 3$ unitary matrix $V$, whose
nine elements are commonly parameterized in terms of three
rotation angles ($\theta^{}_{12}$, $\theta^{}_{23}$,
$\theta^{}_{13}$) and three CP-violating phases ($\delta$, $\rho$,
$\sigma$) \cite{FX01}. The phase parameters $\rho$ and $\sigma$,
which have nothing to do with CP violation in neutrino
oscillations, are usually referred as to the Majorana phases. A
global analysis of current neutrino oscillation data yields
$30^\circ < \theta_{12} < 38^\circ$, $36^\circ < \theta_{23} <
54^\circ$ and $\theta_{13} < 10^\circ$ at the $99\%$ confidence
level \cite{Vissani}, but three phases of $V$ remain entirely
unconstrained. While the absolute mass scale of three neutrinos is
not yet fixed, their two mass-squared differences have already
been determined to a good degree of accuracy \cite{Vissani}:
$\Delta m^2_{21} \equiv m^2_2 - m^2_1 = (7.2 \cdots 8.9) \times
10^{-5} ~{\rm eV}^2$ and $\Delta m^2_{32} \equiv m^2_3 - m^2_2 =
\pm (2.1 \cdots 3.1) \times 10^{-3} ~{\rm eV}^2$. The on-going and
forthcoming neutrino oscillation experiments will shed light on
the sign of $\Delta m^2_{32}$, the magnitude of $\theta^{}_{13}$
and even the CP-violating phase $\delta$.

From a phenomenological point of view, at least two lessons can be
learnt from current experimental data:
\begin{itemize}
\item       The lightest neutrino is allowed to be massless; i.e.,
either $m^{}_1 =0$ (normal neutrino mass hierarchy) or $m^{}_3 =0$
(inverted neutrino mass hierarchy) has no conflict with the
present neutrino oscillation measurements. In both cases, the
non-vanishing neutrino masses can be determined in terms of
$\Delta m^2_{21}$ and $|\Delta m^2_{32}|$:
\begin{eqnarray}
m^{}_1 \; = \; 0 & ~~ \Longrightarrow ~~ & \left \{ \matrix{m^{}_2
\; =\; \sqrt{\Delta m^2_{21}} \; \approx \; 8.94 \times 10^{-3} ~
{\rm eV} \; , \;\;\;\;\;\;\;\;\;\;\;\;\;\;\; \cr m^{}_3 \; =\;
\sqrt{|\Delta m^2_{32}| + \Delta m^2_{21}} \; \approx \; 5.08
\times 10^{-2} ~ {\rm eV} \; ; \cr} \right . \\ \nonumber \\
m^{}_3 \; = \; 0 & ~~ \Longrightarrow ~~ & \left \{ \matrix{m^{}_1
\; =\; \sqrt{|\Delta m^2_{32}| - \Delta m^2_{21}} \; \approx \;
4.92 \times 10^{-2} ~ {\rm eV} \; , \cr m^{}_2 \; =\;
\sqrt{|\Delta m^2_{32}|} \; \approx \; 5.00 \times 10^{-2} ~ {\rm
eV} \; . \;\;\;\;\;\;\;\;\;\;\;\;\; \cr} \right .
\end{eqnarray}
Whether one of the above two neutrino mass spectra is true or
essentially true remains an open question. But we stress that some
interesting neutrino models, such as the minimal seesaw model
\cite{MSM}, are actually able to predict the neutrino mass
spectrum with either $m^{}_1 =0$ or $m^{}_3 =0$.

\item       A special neutrino mixing pattern, the so-called
tri-bimaximal mixing \cite{HPS},
\begin{equation}
V = \left ( \matrix{ 2/\sqrt{6} & 1/\sqrt{3} & 0 \cr -1/\sqrt{6} &
1/\sqrt{3} & 1/\sqrt{2} \cr 1/\sqrt{6} & -1/\sqrt{3} & 1/\sqrt{2}
\cr} \right ) \; ,
\end{equation}
is particularly favored. It yields $\tan\theta^{}_{12} =
1/\sqrt{2}$ (or $\theta_{12} \approx 35.3^\circ$) for the
large-mixing-angle MSW solution \cite{MSW} to the solar neutrino
problem, $\tan\theta^{}_{23} = 1$ (or $\theta_{23} = 45^\circ$)
for the atmospheric neutrino oscillation, and $\theta_{13} = \rho
= \sigma =0^\circ$. As a direct consequence of $\theta_{13} =
0^\circ$, the CP-violating phase $\delta$ is not well defined.
This interesting neutrino mixing pattern is in general expected to
result from an underlying flavor symmetry (e.g., the discrete
$A^{}_4$ \cite{A4}, $S^{}_3$ \cite{S3} or $\mu$-$\tau$ \cite{MT}
symmetry) in the lepton sector. Such a symmetry must be broken
spontaneously or explicitly, in order to account for both the
observed lepton mass spectra and the realistic neutrino mixing
pattern.
\end{itemize}
One purpose of this paper is just to combine both lessons and
reconstruct the simplest neutrino mass texture for either $m^{}_1
=0$ or $m^{}_3 =0$. Looking back to Eqs. (1) and (2), we find that
$m^{}_1 \ll m^{}_2 \ll m^{}_3$ and $m^{}_3 \ll m^{}_1 \approx
m^{}_2$ hold in the $m^{}_1 =0$ and $m^{}_3 =0$ cases,
respectively. This observation implies that a realistic neutrino
mass texture can be established by starting from the symmetry
limit (a) $m^{}_1 = m^{}_2 =0$ and $m^{}_3 \neq 0$ or (b) $m^{}_1
= m^{}_2 \neq 0$ and $m^{}_3 =0$. We shall show that it is
possible to introduce a particular perturbation, which ensures the
resultant neutrino mass matrix $M^{}_\nu$ to reproduce the
tri-bimaximal mixing pattern or its viable variations with all
entries being formed from small integers and their square roots.

The second purpose of this paper is to incorporate the texture of
$M^{}_\nu$ in the minimal Type-II seesaw model \cite{Gu}, an
economical extension of the standard model with only one heavy
right-handed Majorana neutrino $N$ in addition to the $SU(2)^{}_L$
Higgs triplet $\Delta^{}_L$. We shall focus our interest on the
$m^{}_3 =0$ case, so as to obtain a non-vanishing CP-violating
asymmetry in the lepton-number-violating decays of $N$. Such an
asymmetry arises from the interference between the tree-level
amplitude of $N$ decays and the one-loop vertex correction
mediated by $\Delta^{}_L$. Following the idea of baryogenesis via
leptogenesis \cite{FY} and taking account of the flavor-dependent
effects \cite{Flavor}, we shall show that it is possible to
interpret the observed baryon number asymmetry of the Universe
(i.e., $\eta^{}_B = (6.1 \pm 0.2) \times 10^{-10}$ \cite{WMAP})
via thermal leptogenesis in our model, in which only the electron
flavor plays a role in the lepton-to-baryon conversion.

The remaining part of this paper is organized as follows. In
section II we describe a purely phenomenological way to get viable
variations of the tri-bimaximal neutrino mixing pattern from two
simple textures of the neutrino mass matrix $M^{}_\nu$, one with
$m^{}_1 =0$ and the other with $m^{}_3 =0$. Section III is devoted
to incorporating the texture of $M^{}_\nu$ with $m^{}_3 =0$ and a
non-trivial CP-violating phase in the minimal Type-II seesaw
model, and to calculating the flavor-dependent leptogenesis in
order to account for the cosmological baryon number asymmetry
$\eta^{}_B$. A brief summary of our main results is presented in
section IV.

\section{Deviations from Tri-bimaximal neutrino mixing}

Let us work in the basis where the flavor eigenstates of three
charged leptons are identified with their mass eigenstates (i.e.,
the charged-lepton mass matrix $M^{}_l$ is diagonal, real and
positive). Then the mass eigenstates of three neutrinos
($\nu^{}_1$, $\nu^{}_2$, $\nu^{}_3$) are directly linked to their
flavor eigenstates ($\nu^{}_e$, $\nu^{}_\mu$, $\nu^{}_\tau$)
through the neutrino mixing matrix $V$. If $V$ is of the
tri-bimaximal mixing pattern as given in Eq. (3), it can be
decomposed into a product of two Euler rotation matrices: $V =
O^{}_{23} O^{}_{12}$, where
\begin{eqnarray}
O^{}_{12} & = & \left ( \matrix{ \sqrt{2}/\sqrt{2+x^2} &
x/\sqrt{2+x^2} & 0 \cr -x/\sqrt{2+x^2} & \sqrt{2}/\sqrt{2+x^2} & 0
\cr 0 & 0 & 1 \cr} \right ) \; ,
\nonumber \\
O^{}_{23} & = & \left ( \matrix{ 1 & 0 & 0 \cr 0 & 1/\sqrt{2} &
1/\sqrt{2} \cr 0 & -1/\sqrt{2} & 1/\sqrt{2} \cr} \right ) \;
\end{eqnarray}
with $x=1$. Allowing for small deviations of $x$ from unity, we
are then left with some variations of the tri-bimaximal neutrino
mixing pattern which can fit current or future neutrino
oscillation data to a better degree of accuracy. Our strategy of
reconstructing the neutrino mass matrix $M^{}_\nu$ is three-fold:
(1) we take a proper symmetry limit of $M^{}_\nu$, denoted as
$M^{(0)}_\nu$, which can be diagonalized by the orthogonal
transformation $O^{}_{23}$; (2) we introduce a particular
perturbation to $M^{(0)}_\nu$, denoted as $\Delta M^{}_\nu$, which
can be diagonalized by the orthogonal transformation $O^{}_{23}
O^{}_{12}$; (3) we require that $M^{}_\nu = M^{(0)}_\nu + \Delta
M^{}_\nu$ should also be diagonalized by the transformation
$O^{}_{23} O^{}_{12}$. Of course, the texture of $M^{}_\nu$ ought
to guarantee either $m^{}_1 =0$ or $m^{}_3 =0$.

\subsection{Texture of $M^{}_\nu$ with $m^{}_1 =0$}

In the $m^{}_1 =0$ case, we observe from Eq. (1) that $m^{}_2 \ll
m^{}_3$ holds. Hence a reasonable symmetry limit of $M^{}_\nu$ is
expected to be
\begin{equation}
M^{(0)}_\nu \; =\; c \left ( \matrix{ 0 & 0 & 0 \cr 0 & 1 & 1 \cr
0 & 1 & 1 \cr} \right ) \; ,
\end{equation}
where $c$ is assumed to be real and positive. The $S^{}_2$
permutation symmetry in the $(2,3)$ sector of $M^{(0)}_\nu$
assures that this mass matrix can be diagonalized by the
$O^{}_{23}$ transformation:
\begin{equation}
O^T_{23} M^{(0)}_\nu O^{}_{23} \; =\; c \left ( \matrix{ 0 & 0 & 0
\cr 0 & 0 & 0 \cr 0 & 0 & 2 \cr} \right ) \; .
\end{equation}
In other words, $m^{}_3 = 2c$ and $m^{}_2 = m^{}_1 =0$ hold in the
chosen symmetry limit. A non-vanishing value of $m^{}_2$ and a
generalized tri-bimaximal neutrino mixing pattern can result from
the perturbation
\begin{equation}
\Delta M^{}_\nu \; =\; c \varepsilon \left ( \matrix{ x^2 & x & -x
\cr x & 1 & -1 \cr -x & -1 & 1 \cr} \right ) \; ,
\end{equation}
where $\varepsilon$ is a small dimensionless quantity, and $x$ is
a positive number of ${\cal O}(1)$. When $x =1$ holds, $\Delta
M^{}_\nu$ has the $S^{}_2$ permutation symmetry in its $(1,2)$
sector. Given the orthogonal transformations in Eq. (4), the
diagonalization
\begin{equation}
\left ( O^{}_{23} O^{}_{12} \right )^T \Delta M^{}_\nu \left (
O^{}_{23} O^{}_{12} \right ) \; = \; c \varepsilon O^T_{12} \left
( \matrix{ x^2 & \sqrt{2} ~x & 0 \cr \sqrt{2} ~x & 2 & 0 \cr 0 & 0
& 0\cr} \right ) O^{}_{12} \; = \; c \varepsilon \left ( \matrix{
0 & 0 & 0 \cr 0 & 2+x^2 & 0 \cr 0 & 0 & 0 \cr} \right ) \;
\end{equation}
works. The neutrino mass matrix
\begin{equation}
M^{}_\nu \; = \; M^{(0)}_\nu + \Delta M^{}_\nu \; =\; c \left [
\left ( \matrix{ 0 & 0 & 0 \cr 0 & 1 & 1 \cr 0 & 1 & 1 \cr} \right
) + \varepsilon \left ( \matrix{ x^2 & x & -x \cr x & 1 & -1 \cr
-x & -1 & 1 \cr} \right ) \right ] \;
\end{equation}
can then be diagonalized by the unitary matrix $V = O^{}_{23}
O^{}_{12}$:
\begin{equation}
V^T M^{}_\nu V \; =\; O^T_{23} M^{(0)}_\nu O^{}_{23} + \left (
O^{}_{23} O^{}_{12} \right )^T \Delta M^{}_\nu \left ( O^{}_{23}
O^{}_{12} \right ) \; =\; c \left ( \matrix{ 0 & 0 & 0 \cr 0 &
\left (2+x^2 \right ) \varepsilon & 0 \cr 0 & 0 & 2 \cr} \right )
\; .
\end{equation}
Three neutrino mass eigenvalues of $M^{}_\nu$ turn out to be
$m^{}_1 =0$, $m^{}_2 = \left (2+x^2 \right ) c \varepsilon$ and
$m^{}_3 = 2c$. Taking account of Eq. (1), we immediately obtain
the results $c = m^{}_3/2 \approx 2.54 \times 10^{-2}$ eV and
$\varepsilon = 2 m^{}_2/\left [ \left (2+x^2 \right ) m^{}_3
\right ] \approx 0.35/\left (2+x^2 \right )$.

\subsection{Texture of $M^{}_\nu$ with $m^{}_3 =0$}

In the $m^{}_3 =0$ case, we observe from Eq. (2) that $m^{}_1
\approx m^{}_2$ holds. Thus a reasonable symmetry limit of
$M^{}_\nu$ is expected to be
\begin{equation}
M^{(0)}_\nu \; =\; c \left ( \matrix{ 2 & 0 & 0 \cr 0 & 1 & -1 \cr
0 & -1 & 1 \cr} \right ) \; ,
\end{equation}
where $c$ is also assumed to be real and positive. This neutrino
mass matrix can similarly be diagonalized by the $O^{}_{23}$
transformation:
\begin{equation}
O^T_{23} M^{(0)}_\nu O^{}_{23} \; =\; c \left ( \matrix{ 2 & 0 & 0
\cr 0 & 2 & 0 \cr 0 & 0 & 0 \cr} \right ) \; .
\end{equation}
Namely, $m^{}_1 = m^{}_2 = 2c$ and $m^{}_3 =0$ hold in the chosen
symmetry limit. To break the degeneracy of $m^{}_1$ and $m^{}_2$,
we may introduce the same perturbation to $M^{(0)}_\nu$ as that
given in Eq. (7), which can be diagonalized by the same
transformation as that shown in Eq. (8). It is then possible to
diagonalize the neutrino mass matrix
\begin{equation}
M^{}_\nu \; = \; M^{(0)}_\nu + \Delta M^{}_\nu \; =\; c \left [
\left ( \matrix{ 2 & 0 & 0 \cr 0 & 1 & -1 \cr 0 & -1 & 1 \cr}
\right ) + \varepsilon \left ( \matrix{ x^2 & x & -x \cr x & 1 &
-1 \cr -x & -1 & 1 \cr} \right ) \right ] \;
\end{equation}
by using the orthogonal matrix $V = O^{}_{23} O^{}_{12}$:
\begin{equation}
V^T M^{}_\nu V \; =\; O^T_{23} M^{(0)}_\nu O^{}_{23} + \left (
O^{}_{23} O^{}_{12} \right )^T \Delta M^{}_\nu \left ( O^{}_{23}
O^{}_{12} \right ) \; =\; c \left ( \matrix{ 2 & 0 & 0 \cr 0 & 2 +
\left (2+x^2 \right ) \varepsilon & 0 \cr 0 & 0 & 0 \cr} \right )
\; .
\end{equation}
In this case, three neutrino mass eigenvalues of $M^{}_\nu$ are
$m^{}_1 = 2c$, $m^{}_2 = \left [2 + \left ( 2+x^2 \right )
\varepsilon \right ] c$ and $m^{}_3 = 0$. With the help of Eq.
(2), one may easily arrive at $c = m^{}_1/2 \approx 2.46 \times
10^{-2}$ eV and $\varepsilon = 2 \left (m^{}_2 - m^{}_1 \right )
/\left [ \left ( 2+x^2 \right ) m^{}_1 \right ] \approx 3.25
\times 10^{-2}/\left (2+x^2 \right )$.

\subsection{Neutrino mixing patterns}

Although the tri-bimaximal neutrino mixing pattern is of great
interest, it is by no means unique in describing current neutrino
oscillation data. Hence we have gone beyond this pattern by
allowing for $x \neq 1$ in the above discussions. In both case (A)
and case (B), the neutrino mixing matrix $V = O^{}_{23} O^{}_{12}$
reads
\begin{equation}
V^\prime \; =\; \left ( \matrix{ 2/\sqrt{2 \left (2+x^2\right )} &
x/\sqrt{2+x^2} & 0 \cr -x/\sqrt{2\left (2+x^2\right )} &
1/\sqrt{2+x^2} & 1/\sqrt{2} \cr x/\sqrt{2\left (2+x^2\right )} &
-1/\sqrt{2+x^2} & 1/\sqrt{2} \cr} \right ) \; .
\end{equation}
It is obvious that $V$ takes the exact tri-bimaximal mixing
pattern for $x=1$. The allowed range of $x$ can be determined from
that of $\theta^{}_{12}$ through the relationship $x = \sqrt{2}
\tan \theta^{}_{12}$. In view of $30^\circ < \theta^{}_{12} <
38^\circ$, which is obtained from a global analysis of current
neutrino oscillation data \cite{Vissani}, we easily arrive at
$0.82 \lesssim x \lesssim 1.10$. We see that the possibility of $x
= \sqrt{2}$, which leads $V$ to the bimaximal neutrino mixing, has
clearly been excluded. On the other hand, $x=1$ seems to be the
simplest and most favored possibility.

Within the allowed range of $x$, it is not difficult to find out
some viable variations of the tri-bimaximal neutrino mixing
pattern. In particular, we pay interest to such a category of
neutrino mixing matrices $V$: the entries of $V$ are all formed
from small integers and their square roots, which are often
suggestive of a certain flavor symmetry in the language of group
theories. Below are three examples:
\begin{itemize}
\item       $x=\sqrt{6}/3$, corresponding to $\theta^{}_{12} =
30^\circ$ and
\begin{equation}
V \; =\; \left ( \matrix{ \sqrt{3}/2 & 1/2 & 0 \cr -\sqrt{2}/4 &
\sqrt{6}/4 & 1/\sqrt{2} \cr \sqrt{2}/4 & ~~ -\sqrt{6}/4 ~~ &
1/\sqrt{2} \cr} \right ) \; ;
\end{equation}

\item       $x=\sqrt{3}/2$, corresponding to $\theta^{}_{12}
\approx 31.5^\circ$ and
\begin{equation}
V \; =\; \left ( \matrix{ 4/\sqrt{22} & \sqrt{3}/\sqrt{11} & 0 \cr
-\sqrt{3}/\sqrt{22} & 2/\sqrt{11} & 1/\sqrt{2} \cr
\sqrt{3}/\sqrt{22} & -2/\sqrt{11} & 1/\sqrt{2} \cr} \right ) \; ;
\end{equation}

\item       $x=2\sqrt{2}/3$, corresponding to $\theta^{}_{12}
\approx 33.7^\circ$ and
\begin{equation}
V \; =\; \left ( \matrix{ 3/\sqrt{13} & 2/\sqrt{13} & 0 \cr
-\sqrt{2}/\sqrt{13} & 3/\sqrt{26} & 1/\sqrt{2} \cr
\sqrt{2}/\sqrt{13} & -3/\sqrt{26} & 1/\sqrt{2} \cr} \right ) \; .
\end{equation}
\end{itemize}
The pattern of $V$ in Eq. (16) is especially interesting, because
all of its nine elements are formed from four smallest integers
$0$, $1$, $2$, $3$ and their square roots. This pattern has
actually been conjectured in Ref. \cite{Xing03}, but here we
illustrate how it can be obtained from $M^{}_\nu$.

\section{Minimal Type-II seesaw and leptogenesis}

Now let us consider how to derive the neutrino mass matrix
$M^{}_\nu$ in Eq. (9) or Eq. (13) from a specific seesaw model.
One may naively expect that the minimal seesaw model with two
heavy right-handed Majorana neutrinos \cite{MSM} is a good
candidate, because it naturally assures that $M^{}_\nu$ is of rank
2 and has a vanishing mass eigenvalue (either $m^{}_1 =0$ or
$m^{}_3 =0$). Taking account of the fact that $M^{}_\nu$ is
composed of two mass matrices $M^{(0)}_\nu$ and $\Delta M^{}_\nu$,
however, we find that it is more natural to incorporate $M^{}_\nu$
in the minimal Type-II seesaw model with only one heavy
right-handed Majorana neutrino $N$ in addition to the $SU(2)^{}_L$
Higgs triplet \cite{Gu}. In this case, the neutrino mass term can
be written as
\begin{equation}
-{\cal L}^{}_\nu \; = \; \frac{1}{2} \overline{\left (\nu^{~}_L
~N^{\rm c}_R \right )} \left ( \matrix{ M^{}_L & M^{~}_D \cr M^T_D
& M^{}_R \cr} \right ) \left ( \matrix{ \nu^{\rm c}_L \cr N^{}_R
\cr} \right ) ~ + ~ {\rm h.c.} \; ,
\end{equation}
where $\nu^{~}_L$ denotes the column vector of $(\nu^{~}_e,
\nu^{~}_\mu, \nu^{~}_\tau)^{~}_L$ fields, $M^{}_L$ is a $3\times
3$ matrix arising from the leptonic Yukawa interaction induced by
the Higgs triplet $\Delta^{}_L$, $M^{}_D$ is a $3\times 1$ matrix
arising from the leptonic Yukawa interaction induced by the Higgs
doublet $H$, and $M^{}_R = M$ is just the mass of the right-handed
Majorana neutrino $N$. Provided $M$ is considerably higher than
the mass scale of $M^{}_D$, one may obtain the effective
(left-handed) Majorana neutrino mass matrix $M^{}_\nu$ from Eq.
(19) via the well-known Type-II seesaw mechanism \cite{SS2}:
$M^{}_\nu \simeq M^{}_L - M^{}_D M^{-1}_R M^T_D$. Comparing this
formula with $M^{}_\nu = M^{(0)}_\nu + \Delta M^{}_\nu$, we arrive
at
\begin{equation}
M^{(0)}_\nu \; = \; M^{}_L \; , ~~~~ \Delta M^{}_\nu \; = \;
-M^{}_D M^{-1}_R M^T_D \; .
\end{equation}
The texture of $M^{}_L = M^{(0)}_\nu$ given in Eq. (5) or Eq. (11)
may easily be obtained from certain flavor symmetries (such as the
discrete $\mu$-$\tau$ \cite{MT} or $S^{}_2$ \cite{FX} symmetry).
On the other hand, the texture of $\Delta M^{}_\nu$ in Eq. (7) can
be derived from Eq. (20) with a unique form of $M^{}_D$,
\begin{equation}
M^{}_D \; =\; i\sqrt{c\varepsilon M} \left (\matrix{ x \cr 1 \cr
-1 \cr} \right ) \; ,
\end{equation}
together with $M^{}_R =M$. We remark that such a seesaw
realization of the texture of $M^{}_\nu$ does not involve any
parameter fine-tuning or cancellation, and thus it is quite
natural.

More interestingly, the minimal Type-II seesaw model under
consideration can offer a possibility of understanding the
cosmological baryon number asymmetry via thermal leptogenesis
\cite{FY}. For simplicity, we assume the mass of $\Delta^{}_L$ is
much higher than that of $N$ such that the CP-violating asymmetry
in the out-of-equilibrium decays of $N$ is in practice the only
source of leptogenesis. We allow $x$ to be complex in $M^{}_D$ and
its imaginary part is just responsible for CP violation in the
model. In the $m^{}_1 =0$ case, a straightforward analysis shows
that $M^\dagger_D M^{}_L M^{}_D =0$ holds due to the special
textures of $M^{}_L$ in Eq. (5) and $M^{}_D$ in Eq. (21), implying
the absence of CP violation in the decays of $N$. Hence we shall
focus our interest on the $m^{}_3 =0$ case in the following.

\subsection{Neutrino Mixing}

As $x$ is now taken to be a complex parameter, the diagonalization
of $M^{}_\nu$ in Eq. (13) turns out to be quite non-trivial. We
need a unitary matrix $V$ to make the transformation $V^\dagger
M^{}_\nu V^* = {\rm Diag} \{m^{}_1, m^{}_2, 0\}$. We obtain two
non-vanishing mass eigenvalues as
\begin{equation}
m_{1}^{} \; = \; c \sqrt{X-Y} \; , ~~~~ m_{2}^{} \; = \; c
\sqrt{X+Y} \; ,
\end{equation}
where
\begin{eqnarray}
X & = & \frac{1}{2} \left ( |x|^2 + 2 \right )^2
\varepsilon_{}^{2} + \left ( x^2 + {x_{}^{*}}^2 \right )
\varepsilon + 4 \left ( 1 + \varepsilon \right ) \; ,
\nonumber \\
Y & = & \sqrt{X^2 - 4 \left|2 + \left(x^2 + 2 \right) \varepsilon
\right|^2} \; .
\end{eqnarray}
In addition,
\begin{equation}
V \; = \; \left ( \matrix{Z_{1}^{} / \sqrt{|Z_{1}^{}|^2 + 2} &
Z_{2}^{} / \sqrt{|Z_{2}^{}|^2 + 2} & 0 \cr -1 / \sqrt{|Z_{1}^{}|^2
+ 2} & 1/\sqrt{|Z_{2}^{}|^2 + 2} & 1 / \sqrt{2} \cr
1/\sqrt{|Z_{1}^{}|^2 + 2} & -1/\sqrt{|Z_{2}^{}|^2 + 2} &
1/\sqrt{2} \cr} \right ) \left ( \matrix{ e^{i\rho} & 0 & 0 \cr 0
& e^{i\sigma} & 0 \cr 0 & 0 & 1 \cr} \right ) \;
\end{equation}
is just the neutrino mixing matrix, where
\begin{equation}
Z_{1}^{} \; = \; \frac{2Y - T_{1}^{}}{T^{}_{2}} \; , ~~~~ Z_{2}^{}
\; =\; \frac{2Y + T_{1}^{}}{T_{2}^{}}
\end{equation}
with
\begin{eqnarray}
T_{1}^{} & = & \left ( |x|^4 - 4 \right ) \varepsilon_{}^{2} + 2
\left ( x^2 + {x_{}^{*}}^2 \right ) \varepsilon - 8 \varepsilon \;
, \nonumber \\
T_{2}^{} & = & 2 x_{}^{*} \left ( |x|^2 + 2 \right )
\varepsilon_{}^{2} + 2 \left ( x + x_{}^{*} \right ) \varepsilon
\; ;
\end{eqnarray}
and
\begin{eqnarray}
\rho & = & \frac{1}{2} \arg \left [ \left ( 2 + \varepsilon x^2
\right ) {Z_{1}^{*}}^2 - 4 \varepsilon x Z_{1}^{*} +4 \left ( 1 +
\varepsilon \right ) \right ] \; ,
\nonumber \\
\sigma & = & \frac{1}{2} \arg \left [ \left ( 2 + \varepsilon x^2
\right ) {Z_{2}^{*}}^2 + 4 \varepsilon x Z_{2}^{*} +4 \left ( 1 +
\varepsilon \right ) \right ] \; .
\end{eqnarray}
Three neutrino mixing angles are $\theta_{12}^{} = \arctan
[(|Z_{2}^{}| \sqrt{|Z_{1}^{}|^2 + 2})/(|Z_{1}^{}|
\sqrt{|Z_{2}^{}|^2 + 2}) ]$, $\theta_{23}^{} = 45^{\circ}$ and
$\theta^{}_{13} =0^\circ$. Because of $m^{}_3 =0$, only the
difference between $\rho$ and $\sigma$ is a physical Majorana
CP-violating phase. If $x$ is real, then there will be no CP
violation and Eq. (24) will be simplified to Eq. (15).

\subsection{Leptogenesis}

The lepton-number-violating and CP-violating decay of $N$ into a
lepton $l^{}_\alpha$ (for $\alpha = e, \mu, \tau$) and a Higgs
boson $H^{\rm c}$ can occur through both tree-level and one-loop
Feynman diagrams. The latter is indeed the one-loop vertex
correction mediated by the $SU(2)^{}_L$ triplet $\Delta^{}_L$ in
the minimal Type-II seesaw model \cite{Gu}, because the one-loop
vertex correction mediated by $N$ itself is CP-conserving (so is
the self-energy diagram of $N$ decays). As a result, the
CP-violating asymmetry between $N\rightarrow l^{}_\alpha + H^{\rm
c}$ and its CP-conjugate process $N\rightarrow l^{\rm c}_\alpha +
H$ arises from the interference between the tree-level amplitude
and the $\Delta^{}_L$-induced vertex correction. For each lepton
flavor $\alpha$, the corresponding CP-violating asymmetry is given
by \cite{Antusch}
\begin{eqnarray}
\epsilon^{}_{\alpha} & \equiv & \frac{\Gamma (N \rightarrow
l^{}_\alpha + H^{\rm c}) - \Gamma (N \rightarrow l^{\rm c}_\alpha
+ H)}{\displaystyle \sum_\alpha \left[\Gamma (N \rightarrow
l^{}_\alpha + H^{\rm c}) + \Gamma (N \rightarrow
l^{\rm c}_\alpha + H)\right]} \nonumber \\
& \simeq & \frac{3 M}{16 \pi v^2} \cdot \frac{\displaystyle
\sum^{}_{\alpha, \beta} {\rm Im} \left [ \left(M_{D}^{*}
\right)_{\alpha 1}^{} \left(M_{D}^{*} \right)_{\beta 1}^{} \left(
M_{L}^{} \right)_{\alpha\beta}^{} \right ]}{\left( M_{D}^{\dagger}
M_{D}^{} \right)_{11}} \; ,
\end{eqnarray}
where $v \equiv \langle H\rangle \simeq 174$ GeV, and $M$ is the
mass of $N$. Taking account of $M^{}_L = M^{(0)}_\nu$ given in Eq.
(11) and $M^{}_D$ given in Eq. (21), we explicitly obtain
\begin{equation}
\epsilon_{e}^{} \; = \; - \frac{3 M c}{8 \pi v^2} \cdot \frac{{\rm
Im} \left [ \left(x_{}^{*} \right)^2 \right ]}{2 + |x|^2} \; ,
~~~~ \epsilon_{\mu}^{} \; = \; \epsilon_{\tau}^{} \; = \; 0 \; .
\end{equation}
The overall CP-violating asymmetry turns out to be $\epsilon =
\epsilon^{}_e + \epsilon^{}_\mu + \epsilon^{}_\tau =
\epsilon^{}_e$. This interesting result implies that only the
electron flavor contributes to leptogenesis in our model. To be
more specific, we assume that $M$ lies in the region $10^9 ~ {\rm
GeV} \leq M \leq 10^{12} ~ {\rm GeV}$, in which only the
$\tau$-lepton Yukawa coupling is in equilibrium \cite{Flavor}. The
flavor-dependents effects are therefore relevant to thermal
leptogenesis.

The CP-violating asymmetry $\epsilon = \epsilon^{}_e$ can give
rise to a net lepton number asymmetry in the Universe, and this
lepton number asymmetry can partially be converted into a net
baryon number asymmetry due to non-perturbative sphaleron
interactions \cite{Flavor}
\begin{eqnarray}
\eta_B^{} \; \simeq \; -0.96 \times 10^{-2} \sum^{}_\alpha
\epsilon^{}_\alpha \kappa^{}_\alpha \; =\; -0.96 \times 10^{-2}
\epsilon^{}_e \kappa^{}_e \; ,
\end{eqnarray}
where the efficiency factors $\kappa^{}_\alpha$ (for $\alpha = e,
\mu, \tau$) measure the flavor-dependent washout effects
associated with the out-of-equilibrium decays of $N$. To evaluate
the size of $\kappa^{}_e$ in Eq. (30), one may introduce a
parameter $K^{}_e = P^{}_e K$, where $P^{}_e \equiv
|(M^{}_D)^{}_{e1}|^2/(M^\dagger_D M^{}_D)^{}_{11}$ and $K =
\tilde{m}/m^{}_*$ with $\tilde{m} \equiv (M^\dagger_D
M^{}_D)^{}_{11}/M$ being the effective neutrino mass and $m^{}_*
\simeq 1.08 \times 10^{-3} ~ {\rm eV}$ being the equilibrium
neutrino mass \cite{BBP}. Explicitly, $P^{}_e = |x|^2/(2+|x|^2)$
and $\tilde{m} = c \varepsilon (2 + |x|^2)$ hold. Since the
relationship between $\kappa^{}_e$ and $K^{}_e$ is rather
complicated, we do not write it out here but refer the reader to
Ref. \cite{Flavor} for details. We just mention that a numerical
analysis yields $K^{}_e \simeq (0.17 \cdots 1.3)$ and $\kappa^{}_e
\simeq (0.467\cdots 0.64)$ in our model. Hence $\epsilon^{}_e$
should be of ${\cal O}(10^{-7})$ and have a minus sign, in order
to correctly reproduce $\eta^{}_B \sim 6 \times 10^{-10}$.

Let us count the number of free parameters relevant to neutrino
masses, flavor mixing and leptogenesis. They are $c$,
$\varepsilon$, $|x|$, $\arg (x)$ and $M$. On the other hand, we
have four observable quantities $m^{}_1$, $m^{}_2$,
$\theta^{}_{12}$ and $\eta^{}_B$ which depend on the magnitudes of
those five parameters. Although five free parameters cannot be
fully determined from four measured quantities, it is possible to
constrain the former by using the latter. We shall carry out a
numerical calculation and illustrate the viable parameter space of
this minimal Type-II seesaw model in the next subsection.

\subsection{Numerical results}

Given $m^{}_3 =0$, $\theta^{}_{23} = 45^\circ$ and $\theta^{}_{13}
= 0^\circ$ in our model, the inputs of our numerical calculation
include $\Delta m^2_{21} = (7.2 \cdots 8.9) \times 10^{-5} ~{\rm
eV}^2$, $\Delta m^2_{32} = -(2.1 \cdots 3.1) \times 10^{-3} ~{\rm
eV}^2$, $\theta^{}_{12} = (30^\circ \cdots 38^\circ)$
\cite{Vissani} and $\eta^{}_B = (5.9 \cdots 6.3) \times 10^{-10}$
\cite{WMAP}. Since $|x|$ is expected to be of ${\cal O}(1)$, we
typically take $|x| \leq 3$ as a reasonable upper limit. Then it
is straightforward to obtain the allowed ranges of $c$,
$\varepsilon$, $|x|$, $\arg (x)$ and $M$ with the help of those
analytical expressions given in Eqs. (22)--(30). We demonstrate
that this minimal Type-II seesaw model can simultaneously account
for current neutrino oscillation data and the cosmological baryon
number asymmetry. Some results and discussions are in order.
\begin{itemize}
\item       First of all, we find that $c$ gradually ranges
between $2.0 \times 10^{-2} ~ {\rm eV}$ and $2.85 \times 10^{-2} ~
{\rm eV}$. There exist a lower bound on $|x|$ and an upper bound
on $\varepsilon$; namely, $|x|^{}_{\rm min} \simeq 0.82$ and
$\varepsilon^{}_{\rm max} \simeq 1.9 \times 10^{-2}$. The
correlated parameter space of $|x|$ and $\varepsilon$ is shown in
FIG. 1, from which one can get a number of nearly tri-bimaximal
neutrino mixing patterns.

\item       FIG. 2 illustrates the correlated parameter space of
$M$ and $\arg (x)$, which can roughly be understood if one takes
into account $\epsilon^{}_e \propto M \sin [ 2\arg (x)]$ as
indicated in Eq. (29). Neglecting the influence of $|x|$ on
$\epsilon^{}_e$, we find that the lower bound on $M$ comes out
around $\arg (x) \sim 45^\circ$ (or equivalently, $\sin [ 2\arg
(x)] \sim 1$): $M^{}_{\rm min} \simeq 1.3 \times 10^9$ GeV. When
$\arg (x)$ approaches zero, $M$ has to approach infinity in order
to assure $\epsilon^{}_e \sim {\cal O}(10^{-7})$. In our model
with $|x| \sim {\cal O}(1)$, the favored range of $M$ is actually
between $10^9$ GeV and $10^{11}$ GeV.

\item       Note that $|x| =1$ is a particularly interesting
possibility, corresponding to $c \simeq (2.0 \cdots 2.8) \times
10^{-2}$ eV, $\varepsilon \simeq (0.76 \cdots 1.75) \times
10^{-2}$, $\arg (x) \simeq (0.152^\circ \cdots 28.7^\circ)$ and $M
\simeq (4.2 \cdots 1000) \times 10^9$ GeV. In this special case,
the model is simplified to a unique version which only contains
four free parameters and they can all be determined from the
experimental values of $m^{}_1$, $m^{}_2$, $\theta^{}_{12}$ and
$\eta^{}_B$. More accurate data will impose much narrower
constraints on the model parameters $c$, $\varepsilon$, $\arg (x)$
and $M$.
\end{itemize}
Finally, it is worthwhile to point out that our neutrino mixing
pattern is stable against radiative corrections. Running from the
seesaw scale $\mu = M$ down to the electroweak scale $\mu = v$,
$m_{3}^{} =0$ keeps unchanged while other two neutrino masses and
three mixing angles can only receive tiny corrections in the
standard model \cite{RGE}. Although the so-called Dirac
CP-violating phase $\delta$ can be generated together with
$\theta^{}_{13}$ from radiative corrections \cite{Luo}, the
resultant CP-violating effect in neutrino oscillations
(characterized by the Jarlskog invariant of ${\cal O}(10^{-7})$ or
smaller in this model) is too small to be observable.

\section{Summary}

In summary, we have proposed a new category of neutrino mass
ans${\rm\ddot{a}}$tze by starting from a combination of two
phenomenological observations: (1) the lightest neutrino mass
might be zero or vanishingly small, and (2) the neutrino mixing
matrix might be the tri-bimaximal mixing pattern or a pattern
close to it. We have shown that a realistic neutrino mass matrix
$M^{}_\nu$ can be established either in the limit of $m^{}_1 =
m^{}_2 =0$ and $m^{}_3 \neq 0$ or in the limit of $m^{}_1 = m^{}_2
\neq 0$ and $m^{}_3 =0$, corresponding to the possibility of
$m^{}_1 =0$ or $m^{}_3 =0$. In both cases, it is possible to
introduce a particular perturbation which ensures the resultant
neutrino mixing matrix to be the tri-bimaximal mixing pattern or
its viable variations with all entries being formed from small
integers and their square roots. We have incorporated the texture
of $M^{}_\nu$ in the minimal Type-II seesaw model with only one
heavy right-handed Majorana neutrino $N$ in addition to the
$SU(2)^{}_L$ Higgs triplet $\Delta^{}_L$. The $m^{}_3 =0$ case has
been discussed in detail to accommodate CP violation in the
lepton-number-violating decays of $N$. We have demonstrated that
our model can simultaneously interpret current neutrino
oscillation data and the cosmological baryon number asymmetry via
thermal leptogenesis, in which only the electron flavor plays a
role in the lepton-to-baryon conversion.

Finally let us remark that both the neutrino mass spectrum and the
flavor mixing angles are well fixed in the proposed model. It is
therefore easy to test them in the near future, when more accurate
experimental data are available.

\vspace{0.5cm}

{\it Acknowledgments:} This work was supported by the NUS research
grant No WBS: R-144-000-178-112 (A.H.C.) and by the National
Natural Science Foundation of China (Z.Z.X.). Z.Z.X. is indebted
to A.H. Chan and C.H. Oh for warm hospitality at the NUS. S.L. and
Z.Z.X. are grateful to S. Zhou for useful discussions.

\newpage

\begin{figure}
\begin{center}
\vspace{-0.5cm}
\includegraphics[bbllx=2.2cm, bblly=6.0cm, bburx=12.2cm, bbury=16.0cm,%
width=7.4cm, height=7.4cm, angle=0, clip=0]{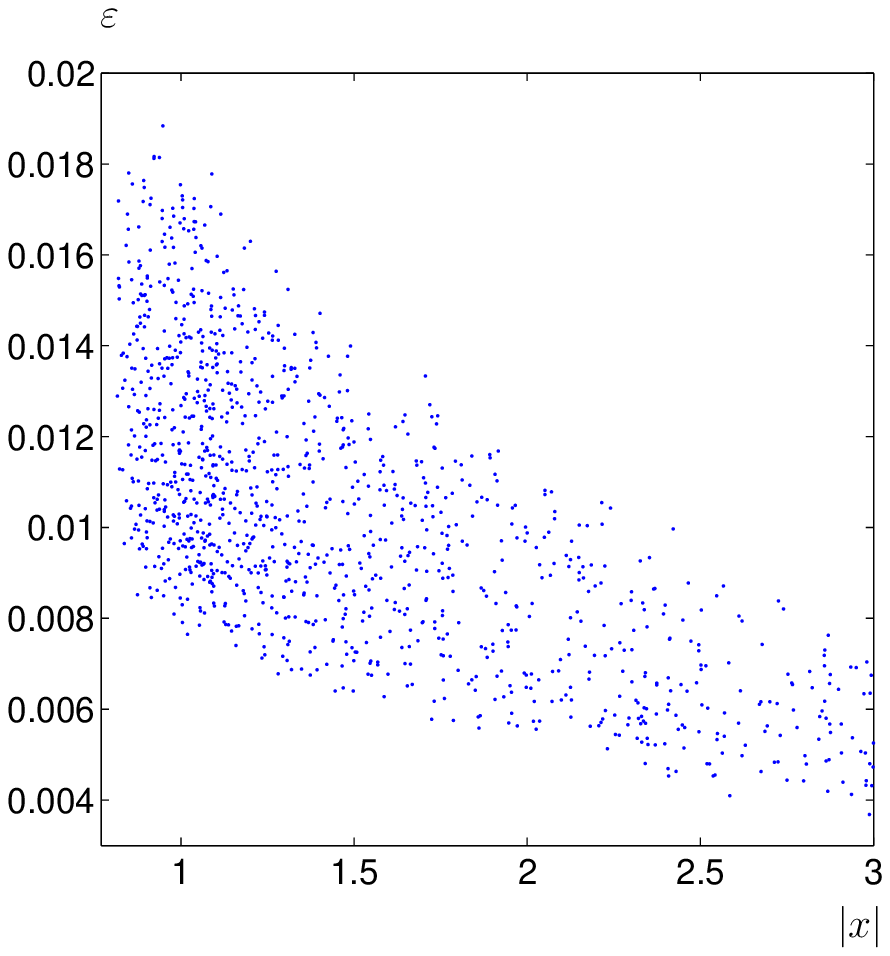}
\vspace{1.8cm}\caption{Parameter space of $|x|$ and $\varepsilon$
constrained by current experimental data on $m^{}_1$, $m^{}_2$,
$\theta^{}_{12}$ and $\eta^{}_B$ in the minimal Type-II seesaw
model with $m^{}_3 =0$.}
\end{center}
\label{fig2}
\end{figure}

\begin{figure}
\begin{center}
\vspace{-0.5cm}
\includegraphics[bbllx=2.2cm, bblly=6.0cm, bburx=12.2cm, bbury=16.0cm,%
width=7.4cm, height=7.4cm, angle=0, clip=0]{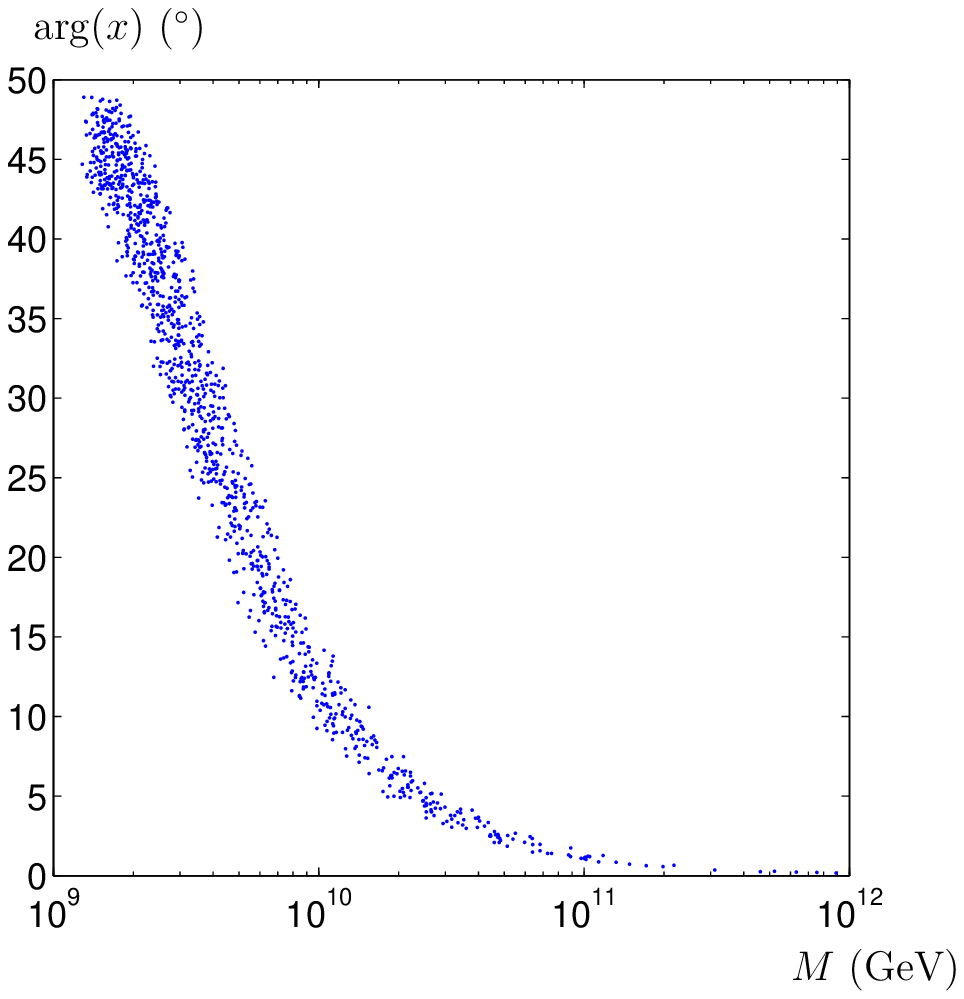}
\vspace{1.8cm}\caption{Parameter space of $M$ and $\arg (x)$
constrained by current experimental data on $m^{}_1$, $m^{}_2$,
$\theta^{}_{12}$ and $\eta^{}_B$ in the minimal Type-II seesaw
model with $m^{}_3 =0$.}
\end{center}
\label{fig3}
\end{figure}


\newpage

\begin{thebibliography}{99}
\bibitem{SNO} SNO Collaboration, Q.R. Ahmad {\it et al.},
Phys. Rev. Lett. {\bf 89}, 011301 (2002).

\bibitem{SK} For a review, see: C.K. Jung {\it et al.},
Ann. Rev. Nucl. Part. Sci. {\bf 51}, 451 (2001).

\bibitem{KM} KamLAND Collaboration, K. Eguchi {\it et al.},
Phys. Rev. Lett. {\bf 90}, 021802 (2003); CHOOZ Collaboration, M.
Apollonio {\it et al.}, Phys. Lett. B {\bf 420}, 397 (1998); Palo
Verde Collaboration, F. Boehm {\it et al.}, Phys. Rev. Lett. {\bf
84}, 3764 (2000).

\bibitem{K2K} K2K Collaboration, M.H. Ahn {\it et al.},
Phys. Rev. Lett. {\bf 90}, 041801 (2003).

\bibitem{FX01} H. Fritzsch and Z.Z. Xing,
Phys. Lett. B {\bf 517}, 363 (2001); Z.Z. Xing, Int. J. Mod. Phys.
A {\bf 19}, 1 (2004).

\bibitem{Vissani} A. Strumia and F. Vissani, Nucl. Phys. B {\bf
726}, 294 (2005); hep-ph/0606054.

\bibitem{MSM} P. Frampton, S.L. Glashow, and T. Yanagida, Phys.
Lett. B {\bf 548}, 119 (2002). For a recent review with extensive
references, see: W.L. Guo, Z.Z. Xing, and S. Zhou, Int. J. Mod.
Phys. E {\bf 16}, 1 (2007).

\bibitem{HPS} P.F. Harrison, D.H. Perkins, and W.G. Scott, Phys.
Lett. B {\bf 530}, 167 (2002); Z.Z. Xing, Phys. Lett. B {\bf 533},
85 (2002); P.F. Harrison and W.G. Scott, Phys. Lett. B {\bf 535},
163 (2002); X.G. He and A. Zee, Phys. Lett. B {\bf 560}, 87
(2003).

\bibitem{MSW} S.P. Mikheyev and A.Yu. Smirnov, Sov. J. Nucl. Phys.
{\bf 42}, 913 (1985); L. Wolfenstein, Phys. Rev. D {\bf 17}, 2369
(1978).

\bibitem{A4} See, e.g., E. Ma, Phys. Rev. D {\bf 70}, 031901 (2004);
New J. Phys. {\bf 6}, 104 (2004); G. Altarelli and F. Feruglio,
Nucl. Phys. B {\bf 720}, 64 (2005); Nucl. Phys. B {\bf 741}, 215
(2006); hep-ph/0610165; K.S. Babu and X.G. He, hep-ph/0507217; A.
Zee, Phys. Lett. B {\bf 630}, 58 (2005); X.G. He, Y.Y. Keum, and
R.R. Volkas, JHEP {\bf 0604}, 039 (2006); E. Ma, H. Sawanaka, and
M. Tanimoto, Phys. Lett. B {\bf 641}, 301 (2006); Y. Koide,
hep-ph/0701018; F. Feruglio, C. Hagedorn, Y. Lin, and L. Merlo,
hep-ph/0702194.

\bibitem{S3} See, e.g., P.F. Harrison and W.G. Scott, Phys. Lett.
B {\bf 557}, 76 (2003); W. Grimus and L. Lavoura, JHEP {\bf 0508},
013 (2005); J.E. Kim and J.C. Park, JHEP {\bf 0605}, 017 (2006);
R.N. Mohapatra, S. Nasri, and H.B. Yu, Phys. Lett. B {\bf 639},
318 (2006); N. Haba, A. Watanabe, and K. Yoshioka, Phys. Rev.
Lett. {\bf 97}, 041601 (2006); Y. Koide, Phys. Rev. D {\bf 73},
057901 (2006); hep-ph/0605074; hep-ph/0612058; S. Kaneko, H.
Sawanaka, T. Shingai, M. Tanimoto, and K. Yoshioka, Prog. Theor.
Phys. {\bf 117}, 161 (2007).

\bibitem{MT} See, e.g.,
T. Fukuyama and H. Nishiura, hep-ph/9702253; R.N. Mohapatra and S.
Nussinov, Phys. Rev. D {\bf 60}, 013002 (1999); Z.Z. Xing, Phys.
Rev. D {\bf 61}, 057301 (2000); Phys. Rev. D {\bf 64}, 093013
(2001); E. Ma and M. Raidal, Phys. Rev. Lett. {\bf 87}, 011802
(2001); C.S. Lam, Phys. Lett. B {\bf 507}, 214 (2001); P.F.
Harrison and W.G. Scott, Phys. Lett. B {\bf 547}, 219 (2002); T.
Ohlsson and G. Seidl, Nucl. Phys. B {\bf 643}, 247 (2002); T.
Kitabayashi and M. Yasu\`{e}, Phys. Rev. D {\bf 67}, 015006
(2003); W. Grimus and L. Lavoura, Phys. Lett. B {\bf 572}, 189
(2003); J. Phys. G {\bf 30}, 73 (2004); Y. Koide, Phys. Rev. D
{\bf 69}, 093001 (2004); R.N. Mohapatra, JHEP {\bf 0410}, 027
(2004); A. de Gouvea, Phys. Rev. D {\bf 69}, 093007 (2004); A.
Ghosal, Mod. Phys. Lett A {\bf 19}, 2579 (2004); W. Grimus, A.S.
Joshipura, S. Kaneko, L. Lavoura, H. Sawanaka, and M. Tanimoto,
Nucl. Phys. B {\bf 713}, 151 (2005); R.N. Mohapatra and W.
Rodejohann, Phys. Rev. D {\bf 72}, 053001 (2005); T. Kitabayashi
and M. Yasu\`{e}, Phys. Lett. B {\bf 621}, 133 (2005); R.N.
Mohapatra, S. Nasri, and H.B. Yu, Phys. Lett. B {\bf 615}, 231
(2005); F. Plentinger and W. Rodejohann, Phys. Lett. B {\bf 625},
264 (2005); I. Aizawa, T. Kitabayashi, and M. Yasu\`{e}, Nucl.
Phys. B {\bf 728}, 220 (2005); K. Matsuda and H. Nishiura, Phys.
Rev. D {\bf 73}, 013008 (2006); Y.H. Ahn, S.K. Kang, C.S. Kim, and
J. Lee, Phys. Rev. D {\bf 73}, 093005 (2006); R.N. Mohapatra, S.
Nasri, and H.B. Yu, Phys. Lett. B {\bf 636}, 114 (2006); Z.Z.
Xing, Phys. Rev. D {\bf 74}, 013010 (2006); K. Fuki and M. Yasue,
Phys. Rev. D {\bf 73}, 055014 (2006). R. Friedberg and T.D. Lee,
HEP$\&$NP {\bf 30}, 591 (2006); Z.Z. Xing, H. Zhang, and S. Zhou,
Phys. Lett. B {\bf 641}, 189 (2006); S. Luo and Z.Z. Xing, Phys.
Lett. B {\bf 646}, 242 (2007); Z.Z. Xing, hep-ph/0703007; Y.H.
Ahn, S.K. Kang, C.S. Kim, and J. Lee, Phys. Rev. D {\bf 75},
013012 (2007).

\bibitem{Gu} P.H. Gu, H. Zhang, and S. Zhou, Phys. Rev. D {\bf
74}, 076002 (2006).

\bibitem{FY} M. Fukugita and T. Yanagida, Phys. Lett. B {\bf 174},
45 (1986).

\bibitem{Flavor} R. Barbieri, P. Creminelli, A. Strumia, and N.
Tetradis, Nucl. Phys. B {\bf 575}, 61 (2000); T. Endoh, T.
Morozumi, and Z.H. Xiong, Prog. Theor. Phys. {\bf 111}, 123
(2004); T. Fujihara, S. Kaneko, S. Kang, D. Kimura, T. Morozumi,
and M. Tanimoto, Phys. Rev. D {\bf 72}, 016006 (2005); A.
Pilaftsis and T.E.J. Underwood, Phys. Rev. D {\bf 72}, 113001
(2005); O. Vives, Phys. Rev. D {\bf 73}, 073006 (2006); A. Abada,
S. Davidson, F.X. Josse-Michaux, M. Losada, and A. Riotto, JCAP
{\bf 0604}, 004 (2006); E. Nardi, Y. Nir, E. Roulet, and J.
Racker, JHEP {\bf 0601}, 164 (2006); A. Abada, S. Davidson, F.X.
Josse-Michaux, M. Losada, and A. Riotto, hep-ph/0605281; Z.Z. Xing
and S. Zhou, hep-ph/0607302; S. Blanchet and P. Di Bari, JCAP {\bf
0606}, 023 (2006); S. Antusch, S.F. King, and A. Riotto, JCAP {\bf
0611}, 011 (2006); S. Pascoli, S.T. Petcov, and A. Riotto,
hep-ph/0609125; G.C. Branco, R. Gonzalez Felipe, and F.R. Joaquim,
Phys. Lett. B {\bf 645}, 432 (2007); S. Antusch and A.M. Teixeira,
JCAP {\bf 0702}, 024 (2007).

\bibitem{WMAP} WMAP Collaboration, D.N. Spergel {\it et al.},
arXiv:astro-ph/0603449.

\bibitem{Xing03} Z.Z. Xing, J. Phys. G {\bf 29}, 2227 (2003);
C. Giunti, hep-ph/0209103.

\bibitem{SS2} R.N. Mohapatra and G. Senjanovic, Phys. Rev. Lett. {\bf 44}, 912
(1980); J. Schechterm and J.W.F. Valle, Phys. Rev. D {\bf 22},
2227 (1980); M. Magg and C. Wetterich, Phys. Lett. B {94}, 61
(1980); G. Lazarides, Q. Shafi and C. Wetterich, Nucl. Phys. B
{\bf 181}, 287 (1981).

\bibitem{FX} H. Fritzsch and Z.Z. Xing, Prog. Part. Nucl. Phys. {\bf 45}, 1
(2000).


\bibitem{Antusch} S. Antusch, arXiv:0704.1591 [hep-ph].

\bibitem{BBP} W. Buchm${\rm \ddot{u}}$ller, P. Di Bari, and M. Pl${\rm
\ddot{u}}$macher, New J. Phys. {\bf 6}, 105 (2004); G.F. Giudice,
A. Notari, M. Raidal, A. Riotto, and A. Strumia, Nucl. Phys. B
{\bf 685}, 89 (2004).

\bibitem{RGE} See, e.g., S. Antusch, J. Kersten, M. Lindner, and M. Ratz,
Nucl. Phys. B {\bf 674}, 401 (2003); J.W. Mei and Z.Z. Xing, Phys.
Rev. D {\bf 69}, 073003 (2004); S. Luo and Z.Z. Xing, Phys. Lett.
B {\bf 632}, 341 (2006).

\bibitem{Luo} See, e.g., S. Luo, J.W. Mei, and Z.Z. Xing, Phys.
Rev. D {\bf 72}, 053014 (2005).
\end{thebibliography}
\end{document}